# Efficient and compact thermo-optic phase shifter in silicon-rich silicon nitride

Hani Nejadriahi,[1] Steve Pappert,[1] Yeshaiahu Fainman[1], Paul Yu[1]

*Department of Electrical & Computer Engineering, University of California, San Diego, 9500 Gilman Drive, La Jolla, CA 92093, USA*
*Corresponding author: hnejadri@eng.ucsd.edu*

**The design, fabrication, and characterization of low loss ultra-compact bends in high index (n = 3.1 at λ = 1550 nm) PECVD silicon rich silicon nitride (SRN) is demonstrated and utilized to realize efficient, small footprint thermo-optic phase shifter. Compact bends are structured into a folded waveguide geometry to form a rectangular spiral within an area of 65 x 65 μm² possessing a total active waveguide length of 1.2 mm. The device features a phase shifting efficiency of 8 mW/π and a 3 dB switching bandwidth of 15 KHz. We propose SRN as a promising thermo-optic platform combining the properties of silicon and silicon stoichiometric nitride.**

____________________________________________________

## 1. Introduction

Phase shifters are important components for many optical signal processing applications, including optical phased arrays (OPA) for beam steering and light detection and ranging (LiDAR) [1]. Such optical systems have a vast set of design requirements [2,3]. Design challenges of photonic integrated circuits (PICs) include maintaining low insertion loss, low power, small footprint, and fast response time while integrating substantial number of components. To meet these performance requirements, it becomes essential to optimally select an appropriate material platform.

For example, high optical power is required in an OPA chip to maximize receiver signal-to-noise ratio. This means a material platform with low two photon absorption (TPA) is ideal. Additionally, to maintain compactness, scalability, high efficiency, and low loss phase shifting, a high index contrast platform with high thermo-optic coefficient is needed.

Silicon-on insulator (SOI) is predominantly used as a CMOS compatible material platform for silicon photonics devices and applications. Due to the high index contrast between silicon and silicon dioxide, SOI enables fabrication of passive components such as compact bends, splitters, and interferometers. Moreover, silicon possesses a high thermo-optic coefficient of (1.86 x 10⁻⁴ K⁻¹) that makes it a viable candidate for thermo-optical phase shifters. However, silicon has its own drawbacks such as a relatively narrower transparency window (1.1 to 8 $\mu m$), TPA, and surface carrier absorption (SCA) leading to low power handling capabilities [2,4,5]. To overcome some of the limitations of silicon, stoichiometric silicon nitride ($Si_3N_4$) is often used as another CMOS compatible platform [6,7]. It has a larger transparency window (0.25 to 8 μm) and possesses negligible TPA unlike silicon. However, it has a lower thermo-optic coefficient (∼ 2 x 10⁻⁵ K⁻¹) and a lower refractive index (∼ 1.98), leading to inefficient optical phase shifters and larger footprint, respectively [2,8]. In light of this, SRN platform is an increasingly attractive CMOS compatible alternative for applications where the benefits of both silicon and stoichiometric silicon nitride are desired. [9,10]. Due to the high degree of tunability of its refractive index and thermo-optic coefficient (n = 3.1 and dn/dT =1.65x10⁻⁴ K⁻¹ at λ = 1550 nm [11]), SRN can be a viable candidate for highly efficient, small footprint devices that require high optical power handling capabilities.

In this manuscript we present means to reduce the power consumption and the footprint of thermo-optic phase shifters by exploiting the design and optimization of ultra-compact bends in SRN. These bends are the structured in a densely folded rectangular spiral of 1.2 mm total length with a total footprint of 65 x 65 μm². For the phase shifter characterization, we use the spiral structure as one of the arms of an unbalanced Mach Zehnder Interferometer (MZI) with serpentine shaped heaters. The experimentally demonstrated device exhibits a $P_\pi$ = 8 mW, a $V_\pi$ = 1.5 V, with extinction ratio exceeding 15 dB proving to be a promising device for thermo-optic applications requiring efficiency, , compactness and high power handling capabilities.

____________________________________________________

## 2. Design modeling and optimization

The thermally induced phase shift for a phase shifter can be represented by:

$$\Delta \phi = \frac{2\pi}{\lambda} \left( \frac{\partial n_{eff}}{\partial T} \right) \Delta T \, L_{mod} \qquad (1)$$

where $\lambda$ is the wavelength of the guided optical mode (in this case 1550 nm), $\frac{\partial n_{eff}}{\partial T}$ is the change in the effective index as a function of temperature change, $L_{mod}$ is the active length of the device and $\Delta T$ is the temperature change across the active length. As seen from the Eq. (1), to achieve efficient phase shifting, a material platform which can provide a large $\frac{\partial n_{eff}}{\partial T}$ is desirable. SRN has this property as its thermo-optic coefficient is close to that of crystalline silicon (1.65 $\pm$ 0.08) x 10⁻⁴ K⁻¹ [11]. In a typical straight waveguide-based phase shifter, increasing the active length of the device will linearly increase the phase change, however at the cost of increased power consumption and slower switching speeds. It would hence be highly desirable to develop ultra-compact bends so as to increase waveguide packing density and achieve a high active length while keeping the heater footprint to a minimum. [12-15]

*A. Compact bend design and simulation results*

It has been previously demonstrated that by varying the precursor gases during plasma enhanced chemical vapor deposition (PECVD), the silicon content and hence the refractive index of SRN can be tuned [11]. In this study, we use one of our highly silicon-rich SRN films with n = 3.1 (at $\lambda$ = 1550 nm) to enable the design of ultra-compact bends. The ultra-compact bends are then used to wrap waveguides in a rectangular spiral fashion for maximizing the length of the waveguide under the metallic heater. The design of 90 and 180-degree bends was hence critical. The bends designed are modified Euler Bends [13-16] and optimized using particle swarm algorithm [17]. Both the 90- and the 180-degree bends are defined by parameters $r_{min}$, $r_{max}$, which are the distances of the inner and outer edge of the waveguide with respect to the bend center and are related by the following relation:

$$r_{max} = r_{min} + f(\theta) \quad (2)$$

$$f(\theta) = w_{max} + (w_{min} - w_{max}) \times \frac{ln(cosh(c \times \theta_0) - c \times \theta)}{ln(cosh(c \times \theta_0))} \quad (3)$$

Where $\theta_0$ is a constant, $\frac{\pi}{4}$ for the 90 degree and $\frac{\pi}{2}$ for the 180-degree bend, θ is the angular position along the bend, $w_{min}$ is the width at the start and end of each bend, $w_{max}$ is the width of the bend at the halfway mark i.e., 45 degrees and 90 degrees for 90 and 180-degree bends respectively. Note the width of the bend at the start and the end is the same and the inlet and outlet waveguides are also kept at the same width, $w_{min}$. The asymmetry constant, c, is determined optimally according to the averaged radius of curvature.

To maintain low loss performance, the width of the waveguide is constantly varied in the bend. This is described by $f(\theta)$. The function $f(\theta)$ is chosen such that the increase in width is very gradual so as to avoid any excess loss as described in Eq. (3). For a given $w_{min}$ optimization of insertion loss was carried out by using particle swarm algorithm and 2.5 D FDTD simulations in LUMERICAL MODE to determine parameters, maximum waveguide widths ($w_{max}$), the constant c, and bend radii, $r_{min}$. The same study is done for the 180-degree bend. The optimized 90-degree bend with an effective bend radius $r_{effective}$ = 4.1 µm and $w_{max}$ = 0.41 was then confirmed to show a low insertion loss of 0.025 dB using full 3D FDTD simulation as shown in Fig.1(a). Fig.1(b), shows the simulation results for a similarly optimized 180-degree bend with an $r_{effective}$ = 4.4 µm and $w_{max}$ = 0.42 µm, showing an insertion loss value of 0.026 dB.

To exploit the small bend radii achieved using the high index SRN and demonstrate efficient thermo-optic phase shifting, we utilized the bends to realize the rectangular spiral shown in Fig.1(c). The spiral of 1.2 mm total propagation length consists of twenty-seven 90-degree bends and two 180-degree bends packed into an area of 65 x 65 µm². Fig.1(d) shows the SEM micrograph of the Y-branch designed and fabricated for TE-polarization with a simulated insertion loss of 0.159 dB. From our previous measurements in [11], the propagation loss for TE polarized optical mode in our SRN waveguides is ~7 dB/cm). While we did not carry out an experimental measurement for the loss in our bends in the complete spiral, the simulated value of loss in our bends can be used to estimate a lower bound for the total insertion loss associated for one such spiral structure to be around~ 1.2 dB.

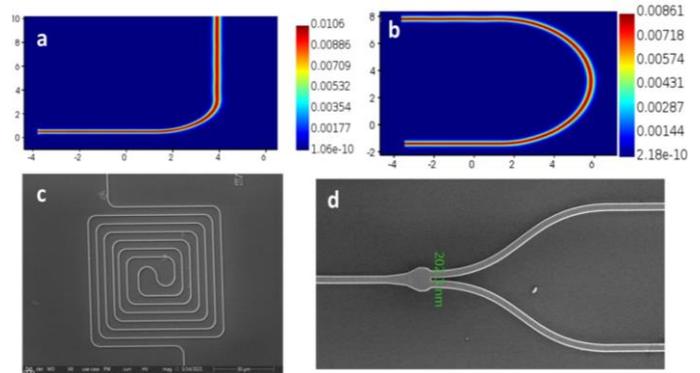

Figure 1. (a) 3D FDTD time averaged electric field intensity profile for the 90-degree bend of $r_{effective}$ = 4.1 µm and $w_{max}$ = 0.41 (b) 3D FDTD time average electric field intensity profile of 180-degree bend with $r_{effective}$ = 4.4 µm and $w_{max}$ = 0.42 µm (c) SEM micrograph of the spiral waveguide of 1.2 mm and 65 x 65 µm² (d) SEM micrograph of the Y-branch coupler used at the input/output of the MZI

*B. Heater geometry and fabrication*

The fabrication of the thermo-optic phase shifters is carried out in a similar fashion to our previous work [11], using the PECVD deposited SRN film of n = 3.1, with the only difference being that in this case, the top cladding thickness is 1 µm.

To study the effect of the metallic heater geometry on the performance of the phase shifter, a serpentine and a rectangular configuration are evaluated. Fig. 2(a) and (b) show the optical microscope (OM) top view of the two configurations integrated onto the spiral phase shifting section of an MZI. The serpentine heater consists of six filaments of 6 µm in width separated by 8 µm; this allows for higher resistance in a compact area, whereas the latter heater is simply a rectangle of 100 x 80 µm². Both configurations have the same metallic (Ni:Cr) thickness of 300 nm. The serpentine heaters were designed in such a way so as to achieve a long metallic filament over the rectangular spiral. The width of the metallic filament and the separation between them were chosen so as to ensure a high value of resistance and fairly uniform temperature profile across all the waveguides while ensuring the ease of fabrication.

Fig. 2(c) shows the distribution of temperature along a cutline passing through the center of waveguides under the metallic heater for the serpentine configuration. Fig. 2(d) shows the corresponding 2D profile of the temperature distribution in a cross section for a heater power of 8 mW. The thermo-optic simulations were carried out using LUMERICAL DEVICE. As expected, waveguides positioned in between the metallic lines (as opposed to directly underneath) experience a slightly lower temperature (97%) than the ones positioned directly underneath the metallic lines [12,18-19].

This ensures temperature uniformity across the spiral waveguide. Similar simulation studies (not shown here) are carried out for the rectangular heater shown in Fig 2(b).

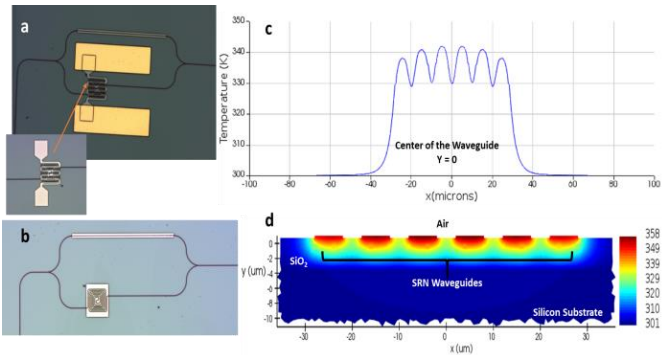

Figure 2. (a) OM top view of the MZI with the spiral as its imbalanced length along with the integrated Ni:Cr serpentine heaters and Cr/Au contact pads -the zoomed-in OM image of the spiral waveguide with the Ni:Cr serpentine heater on top(b) OM top view of the MZI with spiral waveguide and a rectangular Ni:Cr heater on top(c) temperature distribution along a cutline in the plane of the waveguides at y = 0 (d) Simulated temperature profile in a cross-section of the spiral waveguide.

## 3. Results and discussion

To characterize the phase shift as a function of heater power dissipation, the thermo-optic phase shifter is fabricated as part of one arm of an unbalanced MZI. The devices shown in Fig. 2(a, b) are characterized using a fiber-in free space out setup such as that employed in [11]. The estimation of $P_\pi$ was done by operating the device at a fixed wavelength and measuring the output transmission as a function of the power applied to the heater as shown in Fig. 3(a) for a device with a serpentine heater. The power range of 0 to 35 mW corresponds to a voltage applied of up to 4 V, and the extracted $P_\pi$ is 8 ± 0.25 mW ($V_\pi$ = 1.5 ± 0.25 V) with an extinction ratio exceeding 15 dB. The corresponding values for a device with a rectangular heater are $P_\pi \sim$ 11.5± 0.25 mW. Fig. 3(b) shows the spectral shift in the transmission of the MZI as a function of the applied voltage (0 to 1.8 V). The transmission spectra were measured in a range of 1465 nm to 1575 nm (using a CW tunable Agilent laser in the C-Band) showing an FSR of ~ 0.56 nm. The frequency response of the device is characterized by applying a square wave signal (1.5 V amplitude) to the heater such that a $\pi$ phase shift could be achieved. To do so, we use a 15 MHz Hewlett-Packard waveform generator. The frequency of the square wave signal was swept from 0.5 to 25 KHz and the peak-to-peak voltage response was recorded at each step. The 3-dB bandwidth of the phase shifter was measured to be 15±0.5 KHz as shown in Fig.3(c). While the 3-dB is currently low however it should be noted that no optimization has currently been carried out with regards to optimizing the speed of the device. Similar measurements are carried out for the rectangular heater, and no significant changes to the 3-dB cutoff frequency is observed. The uncertainties in the measured values for the $P_\pi$, $V_\pi$ and the 3-dB bandwidth arise from the statistically measured standard deviation of over 15 repeated measurements.

The experimental characteristics of the demonstrated device are promising and can be further improved by optimization of the heater electrode and cladding design. The reduced power consumption of the spiral structure with the serpentine heaters is due to the laterally diffused heat from the heater filaments and the increased fraction of waveguide length per heated area.

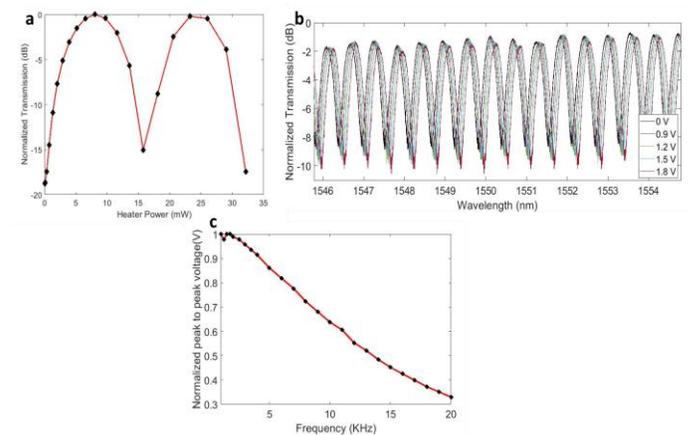

Figure 3. (a) Experimental result on the transmission of the MZI as a function of the switching heater power demonstrating a $P\pi$ = 8 mW. The measurements were carried out at a fixed wavelength, λ = 1551.5 nm to maximize changes in output power as a function of index change (b) Spectral shift of the null point in transmission due to the applied voltage from 0 to 1.8 volts (c) 3-dB cutoff frequency of the device measured to be ~ 15 KHz

**Table 1. Comparison summary of thermo-optic phase shifters**

| Material | $\beta_{TPA}$(cm/GW) (λ = 1550 nm) | Optical Power Handling | $P_\pi$(mW) | Area (mm$^2$) | Ref. |
|---|---|---|---|---|---|
| Silicon (Doped Waveguide) | 0.5 [2] | Low | 24.77 | 0.0015 | 18 |
| | | | 3 | 0.0018 | 16 |
| **SRN (Ni:Cr Heater)** | Negligible [20] | High | 8 | 0.0016 | This work |
| Stoichiometric silicon nitride (TiN Heater) | Negligible [21] | High | 65 | 0.7 | 8 |
| | | | 87.65 | 0.001 | 22 |

The observed performance of the device proves that SRN's high refractive index and high thermo-optic coefficient allow for both compact and efficient phase shifting. Furthermore, SRN phase shifter can be a more desirable option compared to silicon, for applications requiring high optical power handling capabilities and scalability (e.g., OPAs). This is because, as listed in Table 1, SRN (even with a high silicon content and index of up to 3.1) exhibits negligible TPA much lower than that of silicon. Thus, SRN is a great choice for realizing efficient, low footprint, and high optical power handling devices.

## 4. Conclusion

In this work we present the design, fabrication, and characterization of ultra-compact single mode bends using silicon rich nitride. We then use these bends to demonstrate a highly efficient, low footprint SRN thermo-optic phase shifter consisting of a rectangular spiral with integrated serpentine Ni:Cr heaters. The densely folded spiral structure has a 1.2 mm total length with a total footprint of 65 x 65 μm$^2$ consisting of ultra-compact 90 and 180-degree bends with effective radii of 4.1 and 4.4 $\mu m$ respectively. The folded structure allows for an increased overlap of the waveguide with temperature distribution from the integrated serpentine heater and hence a larger optical phase shift to be induced for a given applied thermal power. Two heater geometries were investigated, one a simple rectangle over the spiral waveguide region and the other a serpentine heater. The latter is shown experimentally to outperform the rectangular heater by 30% in terms of its $P_\pi$. The serpentine heater over the phase shifter exhibits a $P_\pi$ = 8 mW and a $V_\pi$ = 1.5 V with an extinction ratio ~ 15 dB. This new thermo-optic phase shifter design enables precise targeting of power dissipation and heat localization, resulting in low thermal crosstalk and high efficiency. The combination of the high index and high thermo-optic coefficient enables not only a compact and efficient phase shifter, but also proves to be the most promising device for applications requiring scalability and high optical power handling capabilities.

**Funding.** National Science Foundation (DMR-1707641, ECCS-1542148); Office of Science (DE-SC0019273); Office of Naval Research (N00014-18-1-2027).

**Acknowledgments.** We thank UCSD's Nano3 cleanroom staff and especially Dr. Maribel Montero, for their assistance in the preparation of the samples.). The SRN synthesis is funded as part of the Quantum Materials for Energy Efficient Neuromorphic Computing, an Energy Frontier Research Center funded by the U.S. Department of Energy, Office of Science, Basic Energy Sciences (DE-SC0019273).

**Disclosures**. The authors declare no conflicts of interest.